\documentclass[11pt]{article}
\usepackage{moriond,epsfig,wrapfig}     

\newcommand{\GeV}{\mbox{\rm ~GeV}~}
\newcommand{\GeVx}{\mbox{\rm ~GeV}}

\newcommand{\GeVsq}{\mbox{${\rm ~GeV}^2$}~}
\newcommand{\GeVsqx}{\mbox{${\rm ~GeV}^2$}}

\newcommand{\pbx}{\mbox{${\rm ~pb}$}}

\newcommand{\pbinv}{\mbox{${\rm ~pb^{-1}}$}}

\newcommand{\ee}{\mbox{$e^+e^-$}}

\newcommand{\lsim}{\raisebox{-0.5mm}{$\stackrel{<}{\scriptstyle{\sim}}$}}

\def\mco{\multicolumn}

\def\be{\begin{equation}}
\def\ee{\end{equation}}
\def\bea{\begin{eqnarray}}
\def\eea{\end{eqnarray}}

\begin{document}
\vspace*{4cm}
\title{Observation of Events with Isolated Charged Leptons and Large Missing Transverse
Momentum and of Events with Multi-Electrons at HERA
}

\author{Tancredi Carli\footnote{
Talk at:
38th Recontres de Moriond Electroweak Interactions and Unified Theories, Les Arc (France) 2003.
}
}

\address{DESY and University of Hamburg, Notkestr. 85, 22607 Hamburg \\
now at: CERN, EP Division, CH-1211 Geneva 23, Switzerland \\
}

\maketitle\abstracts{
Striking events with isolated charged leptons, large missing transverse
momentum and large transverse momentum of the hadronic final state
($P_T^X$) have been observed at the electron proton collider
HERA. 
In the full HERA-I data sample corresponding to an integrated
luminosity of about $130$\pbinv, the H1 experiment observes $10$ events with
isolated electrons or muons and with $P_T^X >25 \GeVx$.
Only $2.9 \pm 0.4$ events are expected from
Standard Model (SM) processes. Six of these
events have $P_T^X >40 \GeVx$, while $1.1 \pm 0.2$ events are
expected. The ZEUS experiment observes good agreement with the SM.
However, in 
a preliminary search ZEUS has found two events with a similar
event topology, but tau-leptons instead of electrons or muons in the final state. 
Only $0.12 \pm 0.02$ events are expected from  SM processes.
Moreover,
six events with two or more electrons forming an invariant
mass bigger than $100$~\GeV have been observed by the H1 experiment.
Three events have two electrons and three events
have three electrons, while only $0.25$ events are expected
in each case.
The ZEUS measurement is in agreement with the SM expectation.
}

\section{Introduction}
Despite the impressive success of the standard 
$SU(3) \otimes SU(2) \otimes U(1)$ 
Model (SM) of particle physics describing the electroweak and
strong interaction between elementary particles in both the
low- and the high-energy regime, it is still unsatisfactory 
in the sense that many
fundamental facts such as the quark-lepton symmetry,
the existence of three generations and their mass spectrum
remain unexplained. Furthermore, the inclusion
of the gravitation as an additional fundamental force in nature
remains an open question.
Recently, hints for the need of an extension of the SM
have been obtained by the observation of neutrino
oscillations suggesting a flavour mixing in the
leptonic sector and non-zero neutrino rest masses \cite{neu_oszis}. 
Only future experiments will be able to fully clarify
the nature of this phenomenon.
An experimental observation of new heavy particles beyond the
presently known particle spectrum would provide an important
guidance toward a deeper understanding of the fundamental particles
and their interactions.

The high centre-of-mass energy
of HERA colliding positrons\footnote{In the data taking period
$1998/1999$ HERA collided electrons on protons. This period
corresponds to an integrated luminosity of $14$\pbinv.} with an energy of
$27.5$\GeV with proton with an energy\footnote{In the data taking period
$1994$-$1997$ HERA collided protons with an energy of $820$\GeVx.
This period corresponds to an integrated luminosity of about $30$\pbinv.}
of $920$~\GeVx, offers the possibility to directly produce 
new heavy particles with a mass up to the centre-of-mass energy of
$\sqrt{s} \approx 320$\GeVx. 
In addition,
the interference of new heavy particles exchanged in the $t$-channel 
with SM processes 
could be experimentally observed as enhancement
or deficit in measured cross-sections over the SM expectation.
In this case HERA is also sensitive to particles with
masses higher than the centre-of-mass energy.

Searches for new particles and new BSM interactions at HERA
have been recently reviewed~\cite{siroiskuze}. Most results obtained
in the first HERA data taking period ($1994$-$2000$) corresponding
to an integrated luminosity of about $130$\pbinv~are covered and
furthermore constrains on various theories
beyond the SM obtained from the HERA-I data are discussed.
Here, the final HERA-I results\cite{h1,ZEUS}  
on the search for isolated electrons or muon
with large missing transverse momentum ($P_T^{\rm miss}$) and
with large transverse momentum of the hadronic final state ($P_T^X$)
are summarised. Furthermore
a preliminary result on the search for isolated tau leptons with
large $P_T^{\rm miss}$ 
is reported\cite{ZEUStau}.  
In addition,
preliminary results on the search for final states with two or three electrons
are presented\cite{H1ee,ZEUSee}.
Searches for other signatures predicted by BSM models are summarised
in another contribution to this conference~\cite{haller}.

\section{Observation of Events with Isolated Leptons and Missing Transverse
Momentum}
\label{sec:isolated}
Large missing transverse momentum, isolated charged leptons and 
a hadronic final state at high transverse energy ($ep \to l\nu X$) 
is a typical possible signature of a singly produced heavy particle
 decaying into a charged lepton and a neutrino. 
Possible new heavy particles postulated by BSM models 
which could be produced at HERA and would 
cause such event signatures have recently been extensively 
discussed\cite{kon,Fritzsch,Rodejohan}.

Within the SM the dominant process leading to 
isolated charged leptons and large $P_T^{\rm miss}$
is the direct production of $W^\pm$-bosons ($ep \to (e) W^\pm X$).
This process has,
however, a relatively small cross-section. 
The inclusive $W^\pm$-boson production cross-section is 
$\sigma(ep \to e W^\pm X) \approx 1 \pbx$. In this process 
the transverse momentum of the hadronic final state $P_T^X$
is expected to be relatively small. If $P_T^X > 25 \GeV$ is required,
the cross-section drops to approximately $0.2$~\pbx.
About two times more $W^+$- than $W^-$-bosons are produced.

\subsection{Calculation of SM Processes for $ep \to (e) W^\pm X$}
\begin{figure}
\begin{center}
\psfig{file=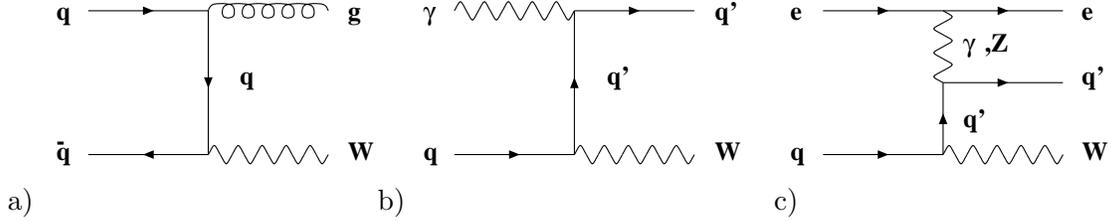,width=14.cm}
\end{center}
\begin{picture}(0,0) \put(10,0){a)} \put(150,0){b)}  \put(300,0){c)}\end{picture}
\vspace{-0.3cm}
\caption{Examples of leading order Feynman diagrams 
for the direct production of $W^\pm$-bosons at HERA. 
\label{fig:w_prod}}
\end{figure}

Typical leading order
Feynman diagrams of this process are shown in Fig.~\ref{fig:w_prod}.
The first two diagrams represent collisions of real photons with
protons (photoproduction). In the second diagram the photon
interacts directly as point-like particle (direct photoproduction), 
in the first diagram it splits into a quark anti-quark pair
before interacting with the hard subprocess (resolved photoproduction). The third
diagrams shows the deep-inelastic scattering (DIS) process, where the
photon is virtual. If the photon virtuality is high, the scattered
electron can be measured in the main detector. 

The direct photoproduction process gives the dominant contribution
to the total cross-section.
The DIS contribution is about a factor of two smaller.
The resolved photoproduction process is a factor of five
smaller at low transverse $W^\pm$-boson momenta and is completely negligible
at high transverse momenta. The process $ep \to  \nu W^\pm X$
only contributes $\approx 5\%$ and can be neglected.

In each of the diagrams the $W^\pm$-boson is radiated from a quark
in the proton.
This is the dominant single contribution of in total six contributing
diagrams (not shown). Of particular interest is
the diagram where the interaction proceeds via the triple gauge boson
coupling ($\gamma W^\pm W^\pm$). This process allows the
anomalous trilinear coupling of gauge bosons to be tested at HERA.

The leading order (${\cal O}(\alpha^2)$) 
calculation of the $ep \to e W^\pm X$ process
has been available as
an event generator since the start-up of HERA\cite{epvec}.
Recently, the QCD corrections ${\cal O}(\alpha^2 \alpha_s)$
have been calculated for the dominant direct photoproduction contribution\cite{dss}.
They include the virtual corrections due to 
one-loop diagrams generated by virtual gluon
exchange and the real correction due to gluon radiation from the
quark lines. In the calculation, the renormalisation and factorisation
scale has been set to $\mu^2 = M_W^2$, where $M_W$ is the mass
of the $W^\pm$-boson. The QCD corrections modify the LO
result by $\pm (10-15\%)$ and reduce the dependence of the
calculated cross-section on  $\mu$ from about $20\%$ to
about  $5 \%$.
Including the uncertainty on the proton parton density function
the total remaining uncertainty on the cross-section prediction 
is about $\pm 5\%$.

\subsection{Observation of Events with Isolated Electrons and Muons}
\label{sec:isolated_emu}
\begin{wraptable}[14]{l}{8.cm}
\vspace{-0.6cm}
\begin{center}
\begin{tabular}{c|c}
H1    & ZEUS \\
\hline
$5^o < \theta_l < 140^o$   & $17^o \lsim \, \theta_l < 115^o$ \\
$P_T^{\rm l} > 10 \GeV$    & $P_T^{\rm l} > 5 \GeV$ \\
$P_T^{\rm miss} > 12 \GeV$ & $P_T^{\rm miss} > 20 \GeV$ \\
$D_{\rm jet} > 1.0$ & $D_{\rm jet} > 1.0$ \\ 
$D_{\rm trk} > 0.5$ & $D_{\rm trk} > 0.5$ \\ 
$\Delta \phi_{lX} < 160^o ($e$), 170^o (\mu)$ & $\Delta \phi_{lX}<172^o$ ($e$)\\
$\delta_{\rm miss} > 5 \GeV$ & $\delta_{\rm miss} > 8 \GeV$ ($e$) \\ 
$V_{ap}/V_{p}<0.5$ ($e$) & \\
$\xi_e^2 > 5000 \GeVsq$ ($e$) & \\
\end{tabular}
\caption{Main H1 and ZEUS event selection cuts for events
with isolated electrons or muon and missing transverse momentum. 
Some of the cuts are only applied in the electron ($e$)
or in the muon ($\mu$) channel. 
\label{tab:iso_cut}}
\end{center}
\end{wraptable}

Events are selected by requiring 
large  $P_T^{\rm miss}$ 
and an electron (e) or muon ($\mu$) in the acceptance
of the calorimeter and/or the tracking system with
high transverse momentum ($P_T^{\rm l}$).
The lepton ($l$) has to be isolated, i.e. the distance in the $\eta-\phi$ 
plane\footnote{The variable $\eta$ denotes the pseudo-rapidity
defined by $\eta=-\ln{\tan{(\theta/2)}}$ and $\phi$ the azimuthal angle.}
from the axis of the closest jet\footnote{Jets are defined by
the inclusive longitudinally invariant $k_T$ algorithm \cite{inclkt} requiring 
a transverse energy $E_T > 5 \GeVx$. ZEUS requires in addition $-1 <\eta<2.5$.}
$D_{\rm jet} = \sqrt{\Delta \eta_{l{\rm jet}}^2 + \Delta \phi_{l{\rm jet}}^2}$
and the distance from the nearest track  
$D_{\rm trk} = \sqrt{\Delta \eta_{l{\rm trk}}^2 + \Delta \phi_{l{\rm trk}}^2}$ has to be large. 
In the electron channel H1 applies the cut on $D_{\rm trk}$ only for electron
polar angles bigger than $45^o$.
To efficiently remove neutral current (NC) DIS events with back-to-back
topologies in the $x$-$y$-plane are rejected by a cut on the
difference between the lepton and the hadronic final state momentum
$\Delta \Phi_{lX}$. Different cut values are applied for the electron and muon 
channel. 
To further reject NC DIS events a cut on 
the longitudinal momentum balance $\delta_{\rm miss}$ is applied.
The variable $\delta_{\rm miss}$ is defined as 
$\delta_{\rm miss}= 2 E_e - \sum_i E_i (1 - \cos{\theta_i})$, where
$E_i$ and $\theta_i$ denotes the energy and the polar angle of each 
energy deposit
and $E_e$ is the electron beam energy. For an event where only
momentum in the proton direction is undetected, e.g. NC DIS events,
$\delta_{\rm miss} =0$.

To gain sensitivity at low $P_T^{\rm miss}$ and to exploit the forward
region of the detector, H1 uses the ratio
of the anti-parallel ($V_{ap}$) to parallel ($V_p$) components of the
transverse momentum measured in the calorimeter. This variable
measures the azimuthal balance of the event. In addition, in the electron channel
the reconstructed squared momentum transfer\footnote{
This observable corresponds to the photon virtuality $Q^2$, if the scattered electron in a
NC DIS process is measured. Since the NC DIS cross-section
steeply falls with $Q^2$, $\xi_e^2$ is generally lower in NC DIS than in the
direct $W^\pm$ process.}
$\xi_e^2=4 E'_e E_e \cos{\theta_e}/2 > 5000 \GeVsqx$, where
$E'_e$ is the energy and $\theta_e$ the polar angle of the
isolated electron, is used for $P_T^X < 25 \GeVx$.
The main event selection criteria for H1 and ZEUS
are summarised in Tab.~\ref{tab:iso_cut}. The main differences
are the requirements on  $P_T^{\rm miss}$ and the lepton acceptance.

These selection criteria are designed to reject the main
background processes with high cross-sections, i.e.
mismeasured NC, $ep \to e X$, 
and charged current (CC), $ep \to \nu X$,  
DIS events, two jet photoproduction events, $ep \to {\rm jet} \, {\rm jet} \, X$,
and events where two leptons are produced in inelastic
photon-photon collisions\footnote{In these processes one of the photons
has high virtuality such that the proton breaks-up. 
See also section \ref{sec:multi}.}, 
$ep \to e l^+ l^- X$.
The H1 event selection has been optimised for direct $W^\pm$-boson
production. H1 achieves a selection efficiency of $40\%$ for
$ep \to (e) W^\pm X$ events with $P_T^X > 25 \GeVx$.
The ZEUS event selection is more oriented toward
the search for a singly produced heavy particle and is less efficient
for $W^\pm$-boson production.

\begin{table}[t]
\begin{center}
\begin{tabular}{c|ccc|ccc}
  \mco{1}{c}{}& \mco{3}{c}{$P_T^X > 25 \GeV$} &  \mco{3}{c}{$P_T^X > 40 \GeV$}  \\
   H1  & Data & SM  & W-contr. & Data & SM  & W-contr. \\
\hline
electron & $4$  &  $1.5 \pm 0.3$ &  $0.8$ & $3$ & $0.5 \pm 0.1$ & $0.45$ \\
muon     & $6$  &  $1.4 \pm 0.3$ &  $1.3$ & $3$ & $0.6 \pm 0.1$ & $0.5$ \\
combined & $10$ &  $2.9 \pm 0.5$ &  $2.6$ & $6$ & $1.1 \pm 0.2$ & $1.0$ \\
\hline
 ZEUS    & Data & SM  & W-contr. & Data & SM  & W-contr. \\
\hline
electron & $2$  &  $2.9 \pm 0.4$ &  $1.3 $ & $0$ & $0.9 \pm 0.1$ & $0.5$ \\
muon     & $5$  &  $2.8 \pm 1.4$ &  $1.4 $ & $0$ & $1.0 \pm 0.1$ & $0.6$ \\
combined & $7$  &  $5.7 \pm 0.6$ &  $2.7 $ & $0$ & $1.9 \pm 0.2$ & $1.1$ \\
\end{tabular}
\caption{Number of measured and expected events in the 
H1 and ZEUS analysis of events with isolated electrons or muons
and large missing transverse momentum. The H1 numbers only include
the $e^+ p$ data sample ($\int {\cal L} dt \approx$ 105 \pbinv). 
The ZEUS numbers correspond to the
full data set ($\int {\cal L} dt \approx$ 130 \pbinv). The number of events expected
by the SM and the one expected only from the $ep \to (e) W^\pm X$ process are given
separately.
\label{tab:iso_h1_results_el_mu}}
\end{center}
\end{table}

In the $e^+ p$ data sample corresponding to an integrated luminosity
of $105 \pbinv$
H1 observes $10$ events in the
electron channel and $8$ events in the muon channel.
An additional prominent event where a scattered muon balances exactly
the transverse momentum of the hadronic system is rejected
by the cut on $\Delta \phi_{lX}$. This spectacular 
event was found in the $1994$
H1 data sample and is discussed elsewhere~\cite{H1_the_event,h1_98,H1susy}. 
In the $e^-p$ data period no event has been found\footnote{H1 has decided to quote
the numbers for the $e^+ p$ and the $e^- p$ region separately.}.

In the electron (muon)
channel $7.2 \pm 1.2$ ($2.23 \pm 0.43$) events are expected from 
$ep \to W^\pm X$ and $2.68 \pm 0.49$ ($0.33 \pm 0.08$) events 
from background processes. One event in the electron channel
has a clearly identified electron ($e^-$), four events
have a positron $e^+$. In the muon channel four events have a
$\mu^+$ and three a $\mu^-$. For the remaining events the charge could
not be determined.

\begin{figure}
\begin{center}
\psfig{figure=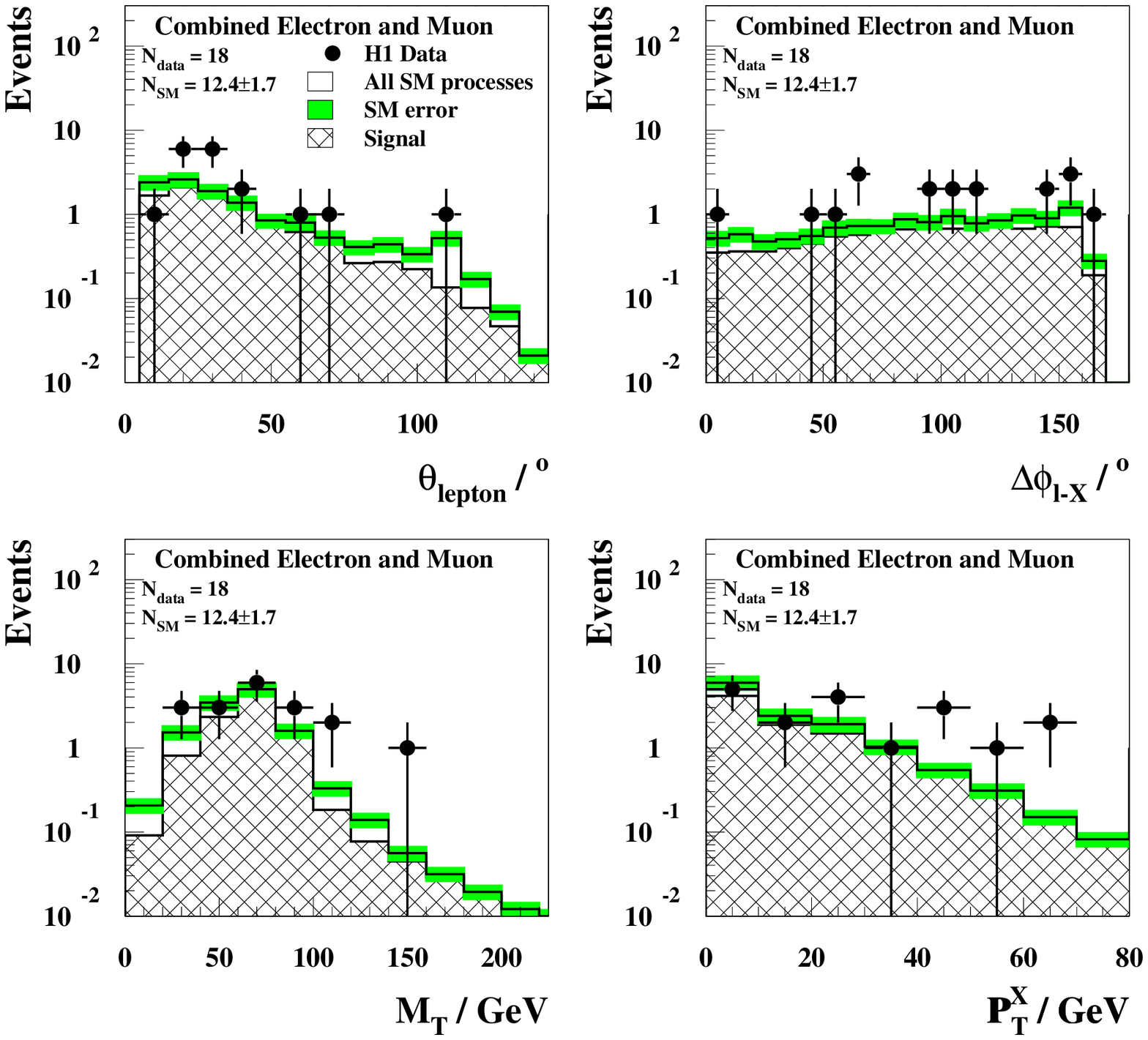,width=16.cm}
\end{center}
\begin{picture}(0,0) 
\put(15,240){a)} \put(250,240){b)}  
\put(15, 45){c)}  \put(250, 45){d)}  
\end{picture}
\vspace{-1.cm}
\caption{Distribution of kinematic variables measured by H1
for events with 
isolated electrons or muons and missing transverse momentum:
a) the polar angle of the lepton, 
b) the azimuthal angle between the lepton and the hadronic
final state,
c) the transverse mass and d) transverse momentum of the
hadronic final state.
The open histogram indicates the expectation for SM processes,
the shaded band the total uncertainty. The hatched
histogram indicates the contribution from direct production of
$W^\pm$-bosons. Also given is the total number of observed
data events ($N_{data}$) and the total number of expected
SM events ($N_{SM}$).
\label{fig:h1_iso_el_mu}}
\end{figure}

The striking feature of these events is their large $P_T^X$. This is shown in 
Fig.~\ref{fig:h1_iso_el_mu}d. While for low $P_T^X$ the number
of measured events roughly corresponds to the number of expected events,
an excess of events is seen toward large $P_T^X$.
The number of measured and expected events with electrons or muons
and  with $P_T^X > 25 \GeV$ or $P_T^X > 40 \GeV$ are summarised in
Tab.~\ref{tab:iso_h1_results_el_mu}.
In the combined electron and muon channel H1 observes $6$ events
with $P_T^X > 40 \GeVx$, while only $1.1 \pm 0.2$ events are expected
from direct $W^\pm$-boson production. 
Other SM processes make only a negligible contribution. 
The Poisson probability that the SM expectation fluctuates
to this observed number of events or more is 
${\cal O}(10^{-3})$.

Some distributions of other variables describing the 
topology of the H1 events are shown in Fig.~\ref{fig:h1_iso_el_mu}.
Most of the events have low 
charged lepton polar angles and are consistent with a uniform
$\Delta \phi_{lX}$ distribution.  
Besides $P_T^X$, most interesting
is the charged lepton-neutrino transverse mass distribution 
defined as
$M_T = \sqrt{{(P_T^{\rm miss} + P_T^l)}^2 -
             {(\vec{P}_T^{\rm miss} + \vec{P}_T^l)}^2}$,
where $\vec{P}_T^{\rm miss}$ and $\vec{P}_T^l$ is the 
$3$-vector of the missing transverse momentum and of the isolated lepton.
For low $M_T$ values the measured events are distributed according
to the Jacobian peak expected from decaying $W^\pm$-boson
and good agreement between the data and the SM expectation
is found. For large $M_T$ values more data events are observed
than expected. Three events have $M_T > 100 \GeVx$. For the
three events where
the scattered electron is detected, the $W^\pm$ mass can be
directly reconstructed. Values consistent with the $W^\pm$-boson are found.


\begin{figure}
\begin{center}
%
\psfig{figure=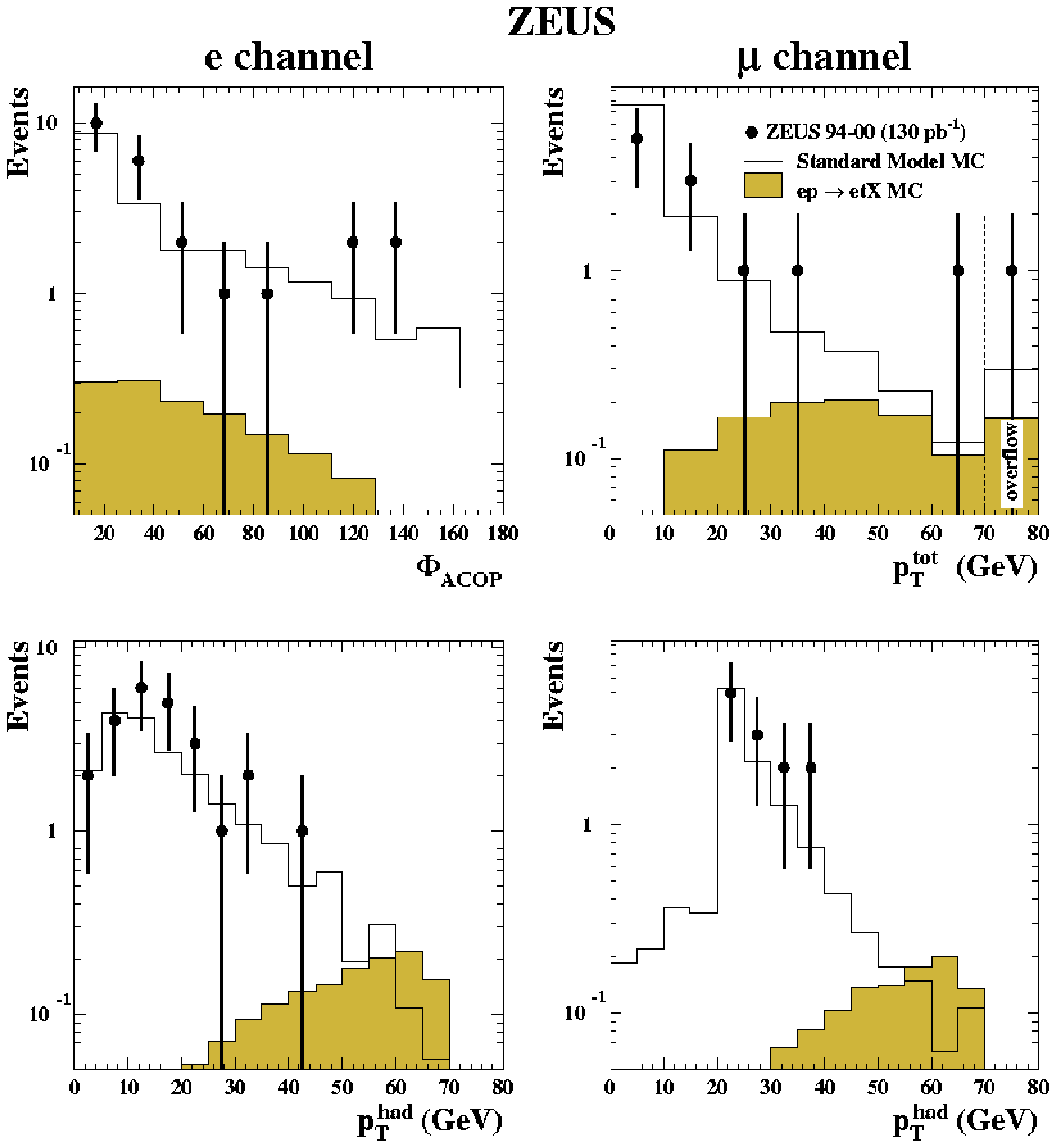,width=16.cm}
\end{center}
\begin{picture}(0,0) 
\put(20,230){a)} \put(250,230){b)}    
\put(20, 25){c)} \put(250, 25){d)}  
\end{picture}
\vspace{-0.5cm}
\caption{Distribution of kinematic variables measured by ZEUS
for events with 
isolated electrons (a, c) or muons (b, d) 
and missing transverse momentum (with relaxed cuts):
a) the azimuthal angle difference between the electron and the hadronic
final state,
b) the total transverse momentum of all particles including the muon,
c) and d) the transverse momentum of the hadronic final state. 
The observable $P_T^{had}$ is identical to $P_T^{X}$ in the text. 
The open histogram shows the SM expectation. The shaded histogram illustrates
a possible contribution of a single top quark produced via an anomalous FCNC
coupling. This histogram is normalised to the upper cross-section limits
obtained by the ZEUS analysis. 
\label{fig:zeus_iso_el_mu}}
\end{figure}

The ZEUS data are in all kinematic
regions consistent with the expected SM background. 
For $P_T^X > 25 \GeV$, $7$ events have been found and $5.7 \pm 0.6$ are expected from SM processes. 
No event with $P_T^X > 40 \GeV$ has been found, but $1.9$ events are
expected.
Fig.~\ref{fig:zeus_iso_el_mu} shows various event topology
variables for the electron (a,c) and the muon (b,d) channel
for a control sample with relaxed cuts.
Only one event in the electron channel and one event in the muon
channel reaches $P_T^X$ values around $40$ \GeVx. 
This event is later rejected, since it does not fulfil the requirement on 
$\delta_{\rm miss}$. 
%
The data are well described by the SM processes 
shown as open histogram. The SM expectation for the electron channel
is dominated by mismeasured NC DIS processes and that
for the muon channel by two photon processes.
The shaded histogram indicates the
expectations from a hypothetical process where a single top quark
$e p \to (e) t X$
is produced via an anomalous flavour changing neutral current
\cite{Fritzsch,singletop}.
None of the measured events has such a topology.

\subsection{Observation of Events with Isolated Tau-Leptons}
\label{sec:isolated_tau}
In view of the excess reported by the H1 collaboration in the electron and muon
channel and not
confirmed by the ZEUS collaboration, it is interesting to 
search for events with isolated tau-leptons and large $P_T^{\rm miss}$
and large $P_T^X$. 
ZEUS has recently presented a preliminary 
analysis \cite{ZEUStau}, where the tau-leptons are identified
in their hadronic decay mode. 
In contrast to jets initiated by
quarks or gluons, jets produced by tau-leptons are
pencil-like, collimated and have a low charged particle multiplicity.

The main background process is CC DIS where a jet from the hadronic final state
is misidentified as tau-lepton.
The CC DIS cross-section is about three orders of magnitude higher
than the signal from direct production of $W^\pm$-bosons followed
by a $W \to \tau \nu_\tau$ decay.
To strongly suppress this large background
while keeping the tau-leptons with a sufficiently large efficiency, 
a multi-variate discrimination technique, 
called PDE-RS\footnote{Probability Density Estimation based on Range
Searching.} was used\cite{PDE-RS}. In this method
the classification of a given event as signal or background 
is based on the signal ($\rho_s$) and the background ($\rho_b$)
probability density 
in the $n$-dimensional vicinity of the event to be classified:
$D = \rho_s/(\rho_s+\rho_b)$.
The probability densities are sampled with MC simulations and
are calculated using a very fast range search algorithm. 

An inclusive CC DIS
Monte Carlo simulation\footnote{Inclusive means here that the event selection is mainly
based on the requirement of large missing transverse energy.
See Ref.~\cite{zeuscc} for details.}
was used as background and a simulation
of direct $W^\pm$-boson production as signal process.
Six observables are exploited to characterise the internal jet structure\footnote{
In general, the internal jet structure is well modeled by the MC
simulations~\cite{jetintstructexp}.}
:
the first and second moment of the radial and longitudinal extension
of the jet energy depositions with respect to the jet axis
\footnote{Jets are defined using the inclusive longitudinally
invariant $K_T$ algorithm \cite{inclkt} and 
$E_T > 5 \GeV$ and $-1 < \eta < 2.5$ is required.},
the subjet multiplicity\footnote{The subjet multiplicity describes
the number of localised energy depositions within a jet that can
be resolved using a certain resolution criterion. An exact definition
can be found in Refs.~\cite{jetintstructexp,jetintstructtheo}.} 
and the observed invariant mass of the particles associated to the jet.

\begin{figure}
\begin{center}
\psfig{figure=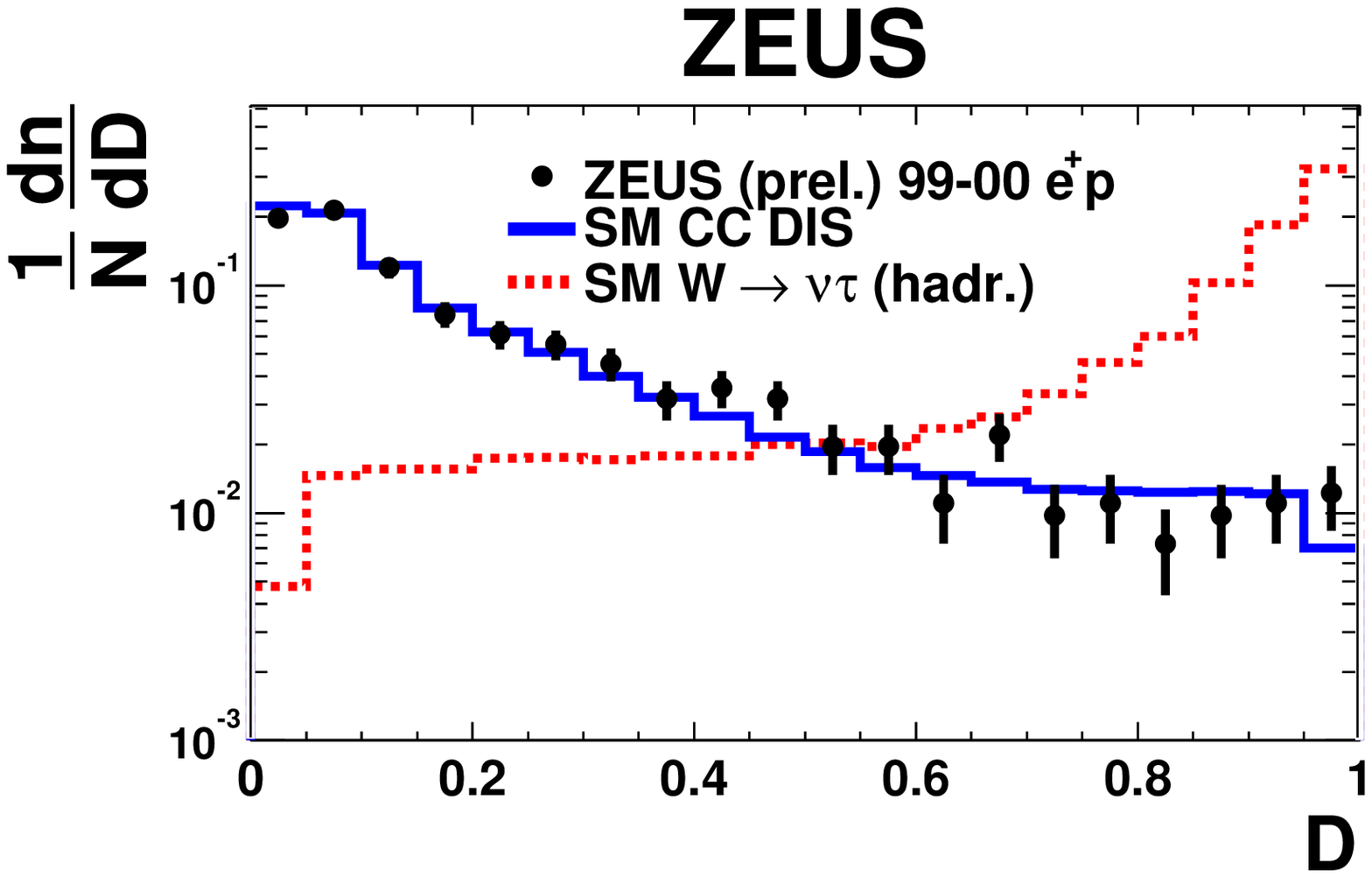,width=10.5cm}
\hspace{-1.1cm}
\psfig{figure=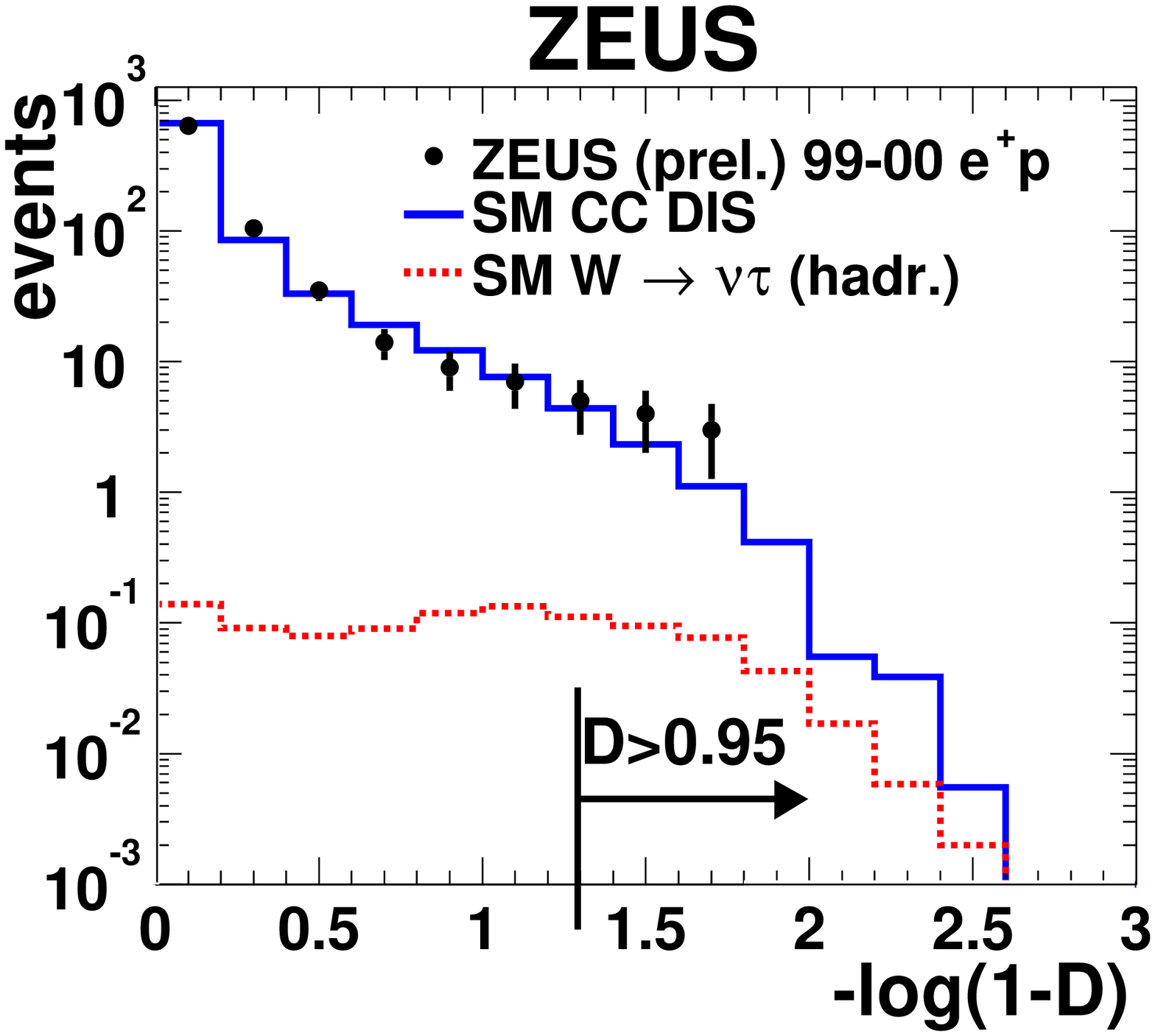,width=6.3cm}
\end{center}
\begin{picture}(0,0) \put(25,20){a)} \put(290,20){b)}  \end{picture}
\vspace{-1.0cm}
\caption{The discriminant $D$ separating hadronic $1$-prong 
tau-lepton decays
from jets induced by quarks or gluons. 
a) shows the shape of the $D$ distribution,
b) shows the absolute number of measured and expected events.
The data measured by ZEUS
correspond to an inclusive selection on missing transverse momentum.
The solid line indicates the background expectation from CC DIS processes, 
the shaded line the signal of tau-leptons
from $W^\pm$-boson decays. 
In b) the $x$-axis is stretched to make the large $D$ values where
the tau-lepton signal is expected more visible.
\label{fig:zeus_tau_discri}}
\end{figure}

Fig.~\ref{fig:zeus_tau_discri}a shows the shape of the
resulting discriminant $D$ for data, background and signal simulation.
The background (signal) is mostly located at low (large) $D$-values.
The shape of the measured discriminant distribution is well described
by the simulation. Only based on the discriminant $D$,
for a signal efficiency of $\epsilon_{\rm sig}=32 \%$, 
a background rejection $1/\epsilon_{\rm bgd}= 154$ is 
obtained.
If in addition jets with only one track are required, the signal
efficiency is still $\epsilon_{\rm sig}=24 \%$ and the
background rejection improves to 
$1/\epsilon_{\rm bgd}= 561$.
The obtained results do not depend on the chosen modeling
of higher order parton radiation (MEPS\footnote{Matrix Element and Parton Shower.} 
or CDM\footnote{Colour Dipole Model.} ).
Using a NC DIS data sample the probability to misidentify
an electron or a jet with one track
as tau-lepton is determined to be 
on the permille level\cite{damir}. The MC simulations predict the
same misidentification probabilities.

In Fig.~\ref{fig:zeus_tau_discri}b the absolute number of 
measured and expected events
is shown. To make the region of large $D$ values more visible 
the $x$-axis is stretched according to $-\log{(1-D)}$.
For this inclusive CC event control 
selection good agreement
between data and simulation is found. However, at the largest
$D$ values slightly more events are found than expected by
the SM processes.


\begin{figure}
\begin{center}
\psfig{figure=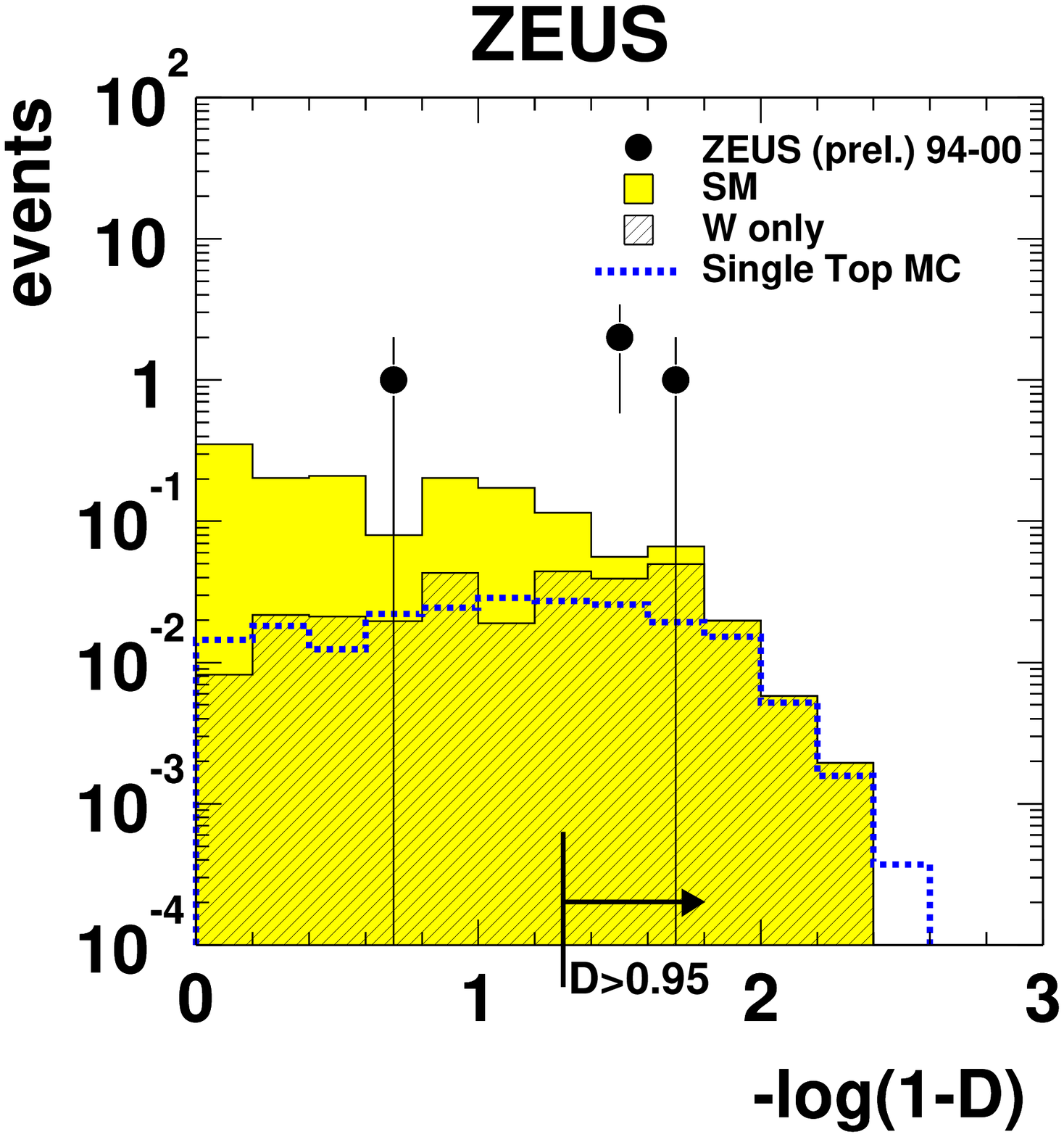,
width=7.cm}
\psfig{figure=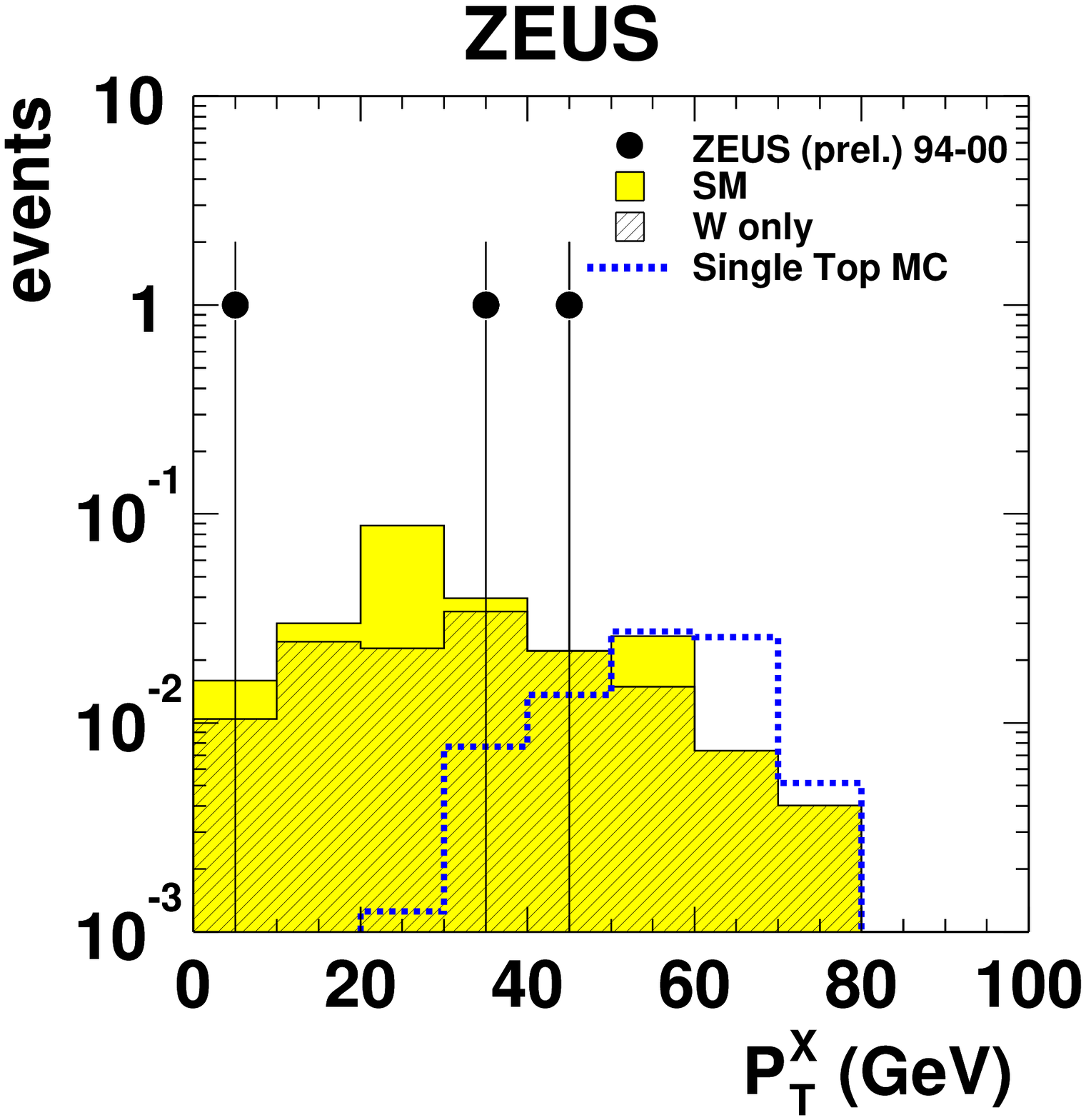,
width=7.cm}
\end{center}
\begin{picture}(0,0) \put(25,20){a)} \put(290,20){b)}  \end{picture}
\vspace{-0.5cm}
\caption{a) The discriminant $D$ for the four events having an isolated
track with high transverse momentum and large missing transverse
energy, b) the $P_T^X$ distribution of the events where a least one
jet is consistent with the tau hypothesis, i.e. $D>0.95$.
\label{fig:zeus_tau-result}}
\end{figure}

When applying similar cuts as in the electron and muon analyses,
but dropping the cut on $P_T^X$ and
the requirement that
the isolated track has to associated with an identified electron
or muon 
and in addition tightening the jet isolation to $D_{jet} > 1.8$,
four events remain in the data sample, while only $1.5 \pm 0.2$
are expected from SM processes. Out of these four events three
are consistent with the tau hypothesis as can be deduced from the
distribution of the tau discriminant (see Fig.~\ref{fig:zeus_tau-result}a).
Out of the three events passing the cut $D>0.95$, two events
have a high transverse momentum, i.e. $P_T^X = 48 \GeV$ and $P_T^X = 37 \GeVx$,
similar to the events found in the
H1 analysis of the electron and the muon channel.

For $P_T^X > 25 \GeVx$ ($P_T^X > 40 \GeVx$)
two (one) events are (is) found in the data and
only $0.12$ (0.06) events are expected from SM processes.
The SM expectation is largely dominated by direct $W^\pm$-boson
production. The Poisson probability to observe two or more events 
when $0.12$ are expected is $6 \cdot 10^{-3}$.


\begin{figure}
\begin{center}
\psfig{figure=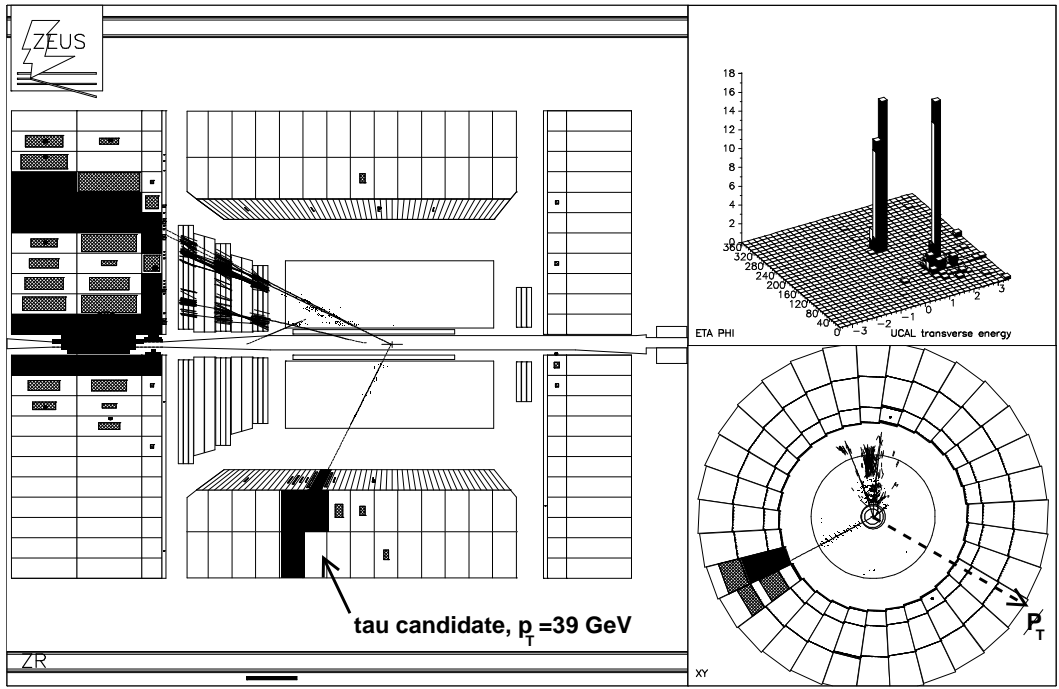,angle=270,
width=14.cm}
\end{center}
\vspace{-0.1cm}
\caption{ZEUS event display of one of the two events with an isolated
tau-lepton and large missing transverse momentum.
\label{fig:zeus_tau_event}}
\end{figure}

\subsection{Conclusions}
The H1 and ZEUS results are summarised in Tab.~\ref{tab:iso_results_all}.
The H1 experiment observes $10$ events in the electron and muon
channel while only $2.9$ events are expected from SM processes.
The excess of measured over expected events has a significance
of approximately three standard deviations. The ZEUS results
are in good agreement with the SM. The effect is therefore
smaller when the two experiments are combined. However, ZEUS
observes $2$ events in the tau-channel where only $0.12$ events
are expected. So far, H1 has not released results in the tau-channel. 

Searches for events where the $W^\pm$-boson decays hadronically
have also been performed\cite{h1,ZEUS}. However, in this
channel the backgrounds are so high that no firm conclusions can be
drawn. In the H1 as well as in the ZEUS analysis the measured
events are in agreement with the SM expectation. Since the
expected number of events from $W^\pm$-boson production is very
small, an anomalous high production cross-section of events with
isolated leptons and missing transverse momentum can not be excluded.

In a random experiment 
the ZEUS measurement in the electron, muon
and tau-lepton channel and the H1 measurement 
would be simulanously produced 
only in about $0.01 \%$ of the cases\cite{dominik}. 
If one allows for an anomalous $W^\pm$ production process,
the ZEUS tau-lepton results would require the largest cross-section 
due to the small selection efficiency. Even in this case
only about in $1\%$ of the cases the H1 and ZEUS measurements
would in a random experiment  be simulanously realised. 
However, the ZEUS tau-lepton results and the H1 results in the electron
and muon channel are compatible within one standard deviation.
If one allows for an anomalous tau-lepton production process
the H1 and ZEUS results would be realised in a few percent of the
cases in a random experiment, i.e. all
the measured results would be consistent.
In this case only tau-leptons are produced and
the electrons and muons signals are generated via the tau-decay.
This assumption also models a coupling which is much stronger for the tau-lepton
than for the leptons of the first two families.
Such an interaction can be
incorporated in supersymmetric models where the R-parity is broken.
Also the production of two heavy Majorana neutrinos $N$ exchanged in the t-channel,
i.e. $ep \to W N W \to \nu l^+ l^+ X$, predicts an enhancement of tau-leptons 
over electrons or muons\footnote{One of the leptons is in most case outside
the detector acceptance.} \cite{Rodejohan}.

The high luminosity expected for the HERA-II data taking period
will be needed to clarify whether the excess of measured events in the
electron, muon and tau-lepton channel over the SM process
are due to a statistical fluctuation or to a new BSM interaction.
Another possible explanation is that the hard spectrum of the hadronic
final state is generated by higher order QCD effects which are presently only 
included to the first order of the strong coupling constant.

Assuming that the ZEUS result in the muon and the electron channel is
a downwards fluctuations and H1 measures the real rate, an additional
data set of $200$\pbinv~will be required to get
an excess of two standard deviations in the ZEUS analysis. 
In this case the H1 excess will be on the five sigma level.
If one assumes that the H1 and ZEUS experiments measured the average rate
in $1994-2000$, an 
additional data set of $500$\pbinv~will be required to observe a deviation
of five standard deviation in the combined H1 and ZEUS analysis. 
The signficance will be 
increased, if the tau-lepton channel is also considered.

\begin{table}[t]
\begin{center}
\begin{tabular}{c|cc|cc|cc|}
 & \mco{4}{c|}{electron/muon channel combined} &  \mco{2}{c|}{tau channel } \\
 &  \mco{2}{c|}{H1} &  \mco{2}{c|}{ZEUS} &   \mco{2}{c|}{ZEUS} \\
 & Data & SM & Data & SM & Data & SM \\
 $P_T^X > 25 \GeV$ &  $10$ &  $2.9 \pm 0.5$ & $7$  &  $5.7 \pm 0.6$ & $2$ & $0.12 \pm 0.02$ \\
 $P_T^X > 40 \GeV$ &  $ 6$ &  $1.1 \pm 0.2$ & $0$  &  $1.9 \pm 0.2$ & $1$ & $0.06 \pm 0.01$ \\
\end{tabular}
\caption{Overview of the results in the combined electron and muon channel
and in the tau-channel.
\label{tab:iso_results_all}}
\end{center}
\end{table}
\newpage
\begin{wraptable}[13]{l}{9.3cm}
\vspace{-0.8cm}
\begin{tabular}{ccccc}
 \mco{5}{l}{$M_{1,2}> 100 \GeVx:$} \\
  &  \mco{4}{c}{H1} \\
 & Data & SM & $\gamma \gamma$ & fake  \\
 $= 2   e^\pm$ &  $ 3$ &  $0.25 \pm 0.05$ & $0.21 \pm 0.04$ & $0.04 \pm 0.03$ \\
 $= 3   e^\pm$ &  $ 3$ &  $0.23 \pm 0.04$ & $0.23 \pm 0.04$ & $0.00 \pm 0.00$ \\
& \mco{4}{c}{ZEUS} \\
& Data & SM & $\gamma \gamma$ & fake  \\
 $= 2   e^\pm$ &  $2$  &  $0.77 \pm 0.08$ & $0.47 \pm 0.05$ & $0.3 \pm 0.09$ \\
 $= 3   e^\pm$ &  $0$  &  $0.37 \pm 0.04$ & $0.37 \pm 0.04$ & $0.0 \pm 0.00$ \\
 $\ge 2 e^\pm$ &  $2$  &  $1.14 \pm 0.09$ & $0.84 \pm 0.06$ & $0.3 \pm 0.09$ \\
\end{tabular}
\caption{Number of events with two, three or two or more
electrons measured by H1 and ZEUS and expected from SM processes. 
\label{tab:sum_multiel}}
\end{wraptable}

\section{Observation of Events with Multi-Electrons at High Transverse Energy}
\label{sec:multi}
The high centre-of-mass energy of HERA offers the possibility of producing
events with two or more leptons at high transverse energy.
The production rates of such events can be accurately predicted
by the SM. These processes are therefore sensitive to
possible small signals beyond the SM.

Events with two electrons are produced in photon-photon ($\gamma-\gamma$)
collisions. The photons are radiated off from the incoming 
proton and the electron, i.e.
$ep \to \gamma \gamma \to e^+ e^- (e^\pm) X$.
Since the cross-section is anti-proportional to the
virtualities $t_1^2$ and $t_2^2$ of the two photons, i.e. 
$d\sigma/dQ^2 \propto \frac{1}{t_1^2} \frac{1}{t_2^2}$,
the photons are real in most cases. However, in some cases one
of the photon virtualities can be larger. Then also the scattered
electron can be measured in the detector. 
The process $\gamma \gamma \to e^+ e^-$ can be simulated
using GRAPE\cite{grape}. It is based on the exact
electroweak matrix element at tree level for $\gamma$ (Z)-$\gamma$ (Z) 
collisions and internal photon conversions. Depending
on the photon virtuality on the proton side
either elastic, quasi-elastic or
inelastic processes are simulated.

Events are selected by requiring an isolated collimated electromagnetic
cluster with high energy and a track associated to it.
H1 requires two electrons in the acceptance of
the central drift chambers ($20^o < \theta < 150^o$) with
a transverse energy $E_T > 5 \GeVx$. The electron with the highest
$E_T$ has to fulfil $E_T > 10 \GeVx$. Additional electrons
can 
have any $E_T$ and do not need to be associated to a track.
ZEUS required two
electrons within the acceptance of the central drift chamber,
i.e. $17.2^o < \theta < 164^o$, and a transverse energy $E_T > 10$ \GeVx.
A track with a momentum well matched to the measured energy has
to be associated with the cluster. Electrons in the forward
and backward regions were measured without a track requirement.
Under certain kinematic conditions they could be used in the
determination of the invariant two electron mass.

One of the
difficulties in this analysis is to control the background
from photons and jets which are misidentified as electrons.
This is particularly difficult in the forward region.
Photons produced in the reaction $ ep \to \gamma e \to \gamma e$ (QED Compton)
can convert near the beam pipe. If the photon conversion
is asymmetric, e.g. the created electron has high momentum and the
created positrons has a low one, the photon can not be distinguished
from a genuine electron. Another background is due to NC DIS 
events where the scattered quark produces a jet with only
one track and a high electromagnetic fraction.

\begin{figure}
\begin{center}
\psfig{figure=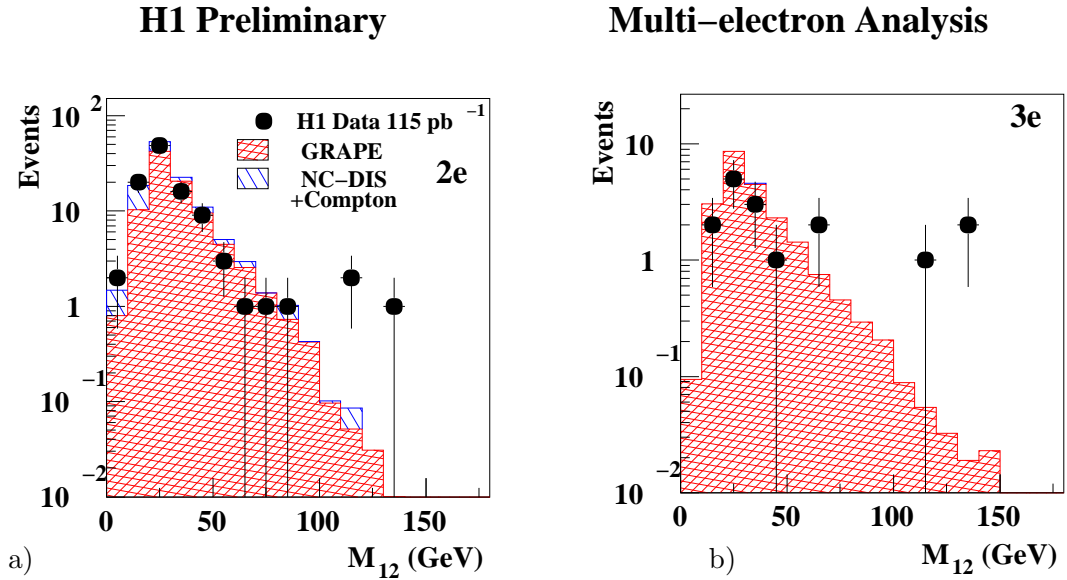,width=14.cm}
\end{center}
\begin{picture}(0,0) \put(25,20){a)} \put(290,20){b)}  \end{picture}
\vspace{-0.7cm}
\caption{
Invariant mass of events with $2$ (a) and
$3$ (b) electrons measured by H1 and expected by SM processes.
\label{fig:h1_multi_electron}}
\end{figure}

Fig.~\ref{fig:h1_multi_electron} shows the invariant mass
distribution ($M_{1,2}$) 
measured by H1 for two (a) and three (b) electron events in a 
data sample collected from $1994$-$2000$
corresponding to an integrated luminosity of $115$ \pbinv.
In the three electron case $M_{1,2}$
is calculated from the electrons with the highest $E_T$.
Shown as histogram is the expectation
of SM processes, i.e. the genuine multi-electron events
and the background from misidentified electrons.
The good agreement between data and SM expectation demonstrates
the good understanding of the $\gamma \gamma \to e^+ e^-$ reaction
and of detector effects. At high masses more events
are measured than expected by the SM processes.
H1 measures three events with two electrons and three events
with three electrons, while only $0.25$ events are expected
in each case.

Also events with two muons instead of two electrons haved analysed.
No excess of the data over the SM expectation is found in the H1 and ZEUS
analyses. However,
due to the smaller available data sample and the lower detection efficiency
this does not contradict
a possible signal in the electron channel.

\begin{figure}
\begin{center}
\psfig{figure=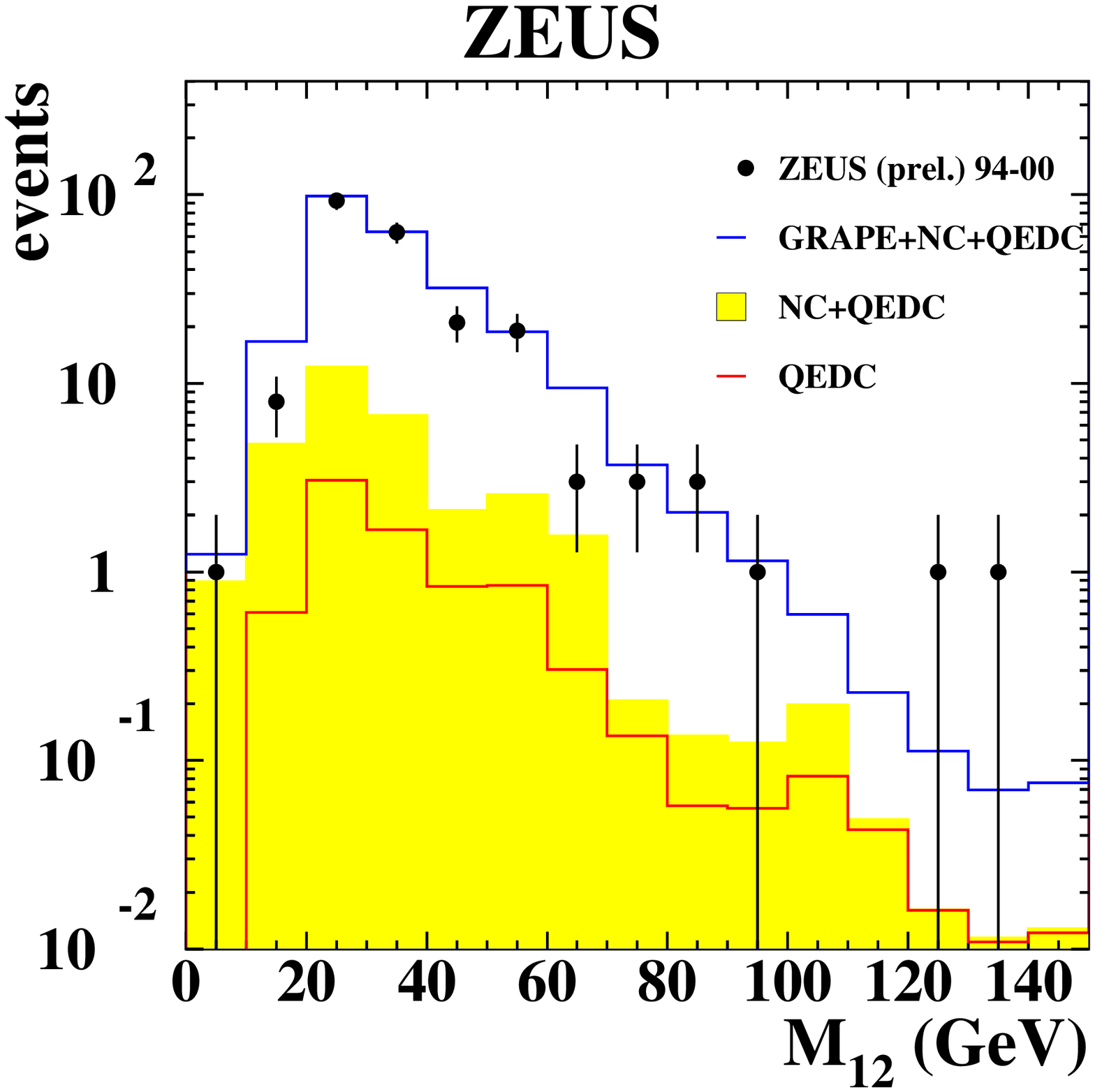,width=7.5cm}
\psfig{figure=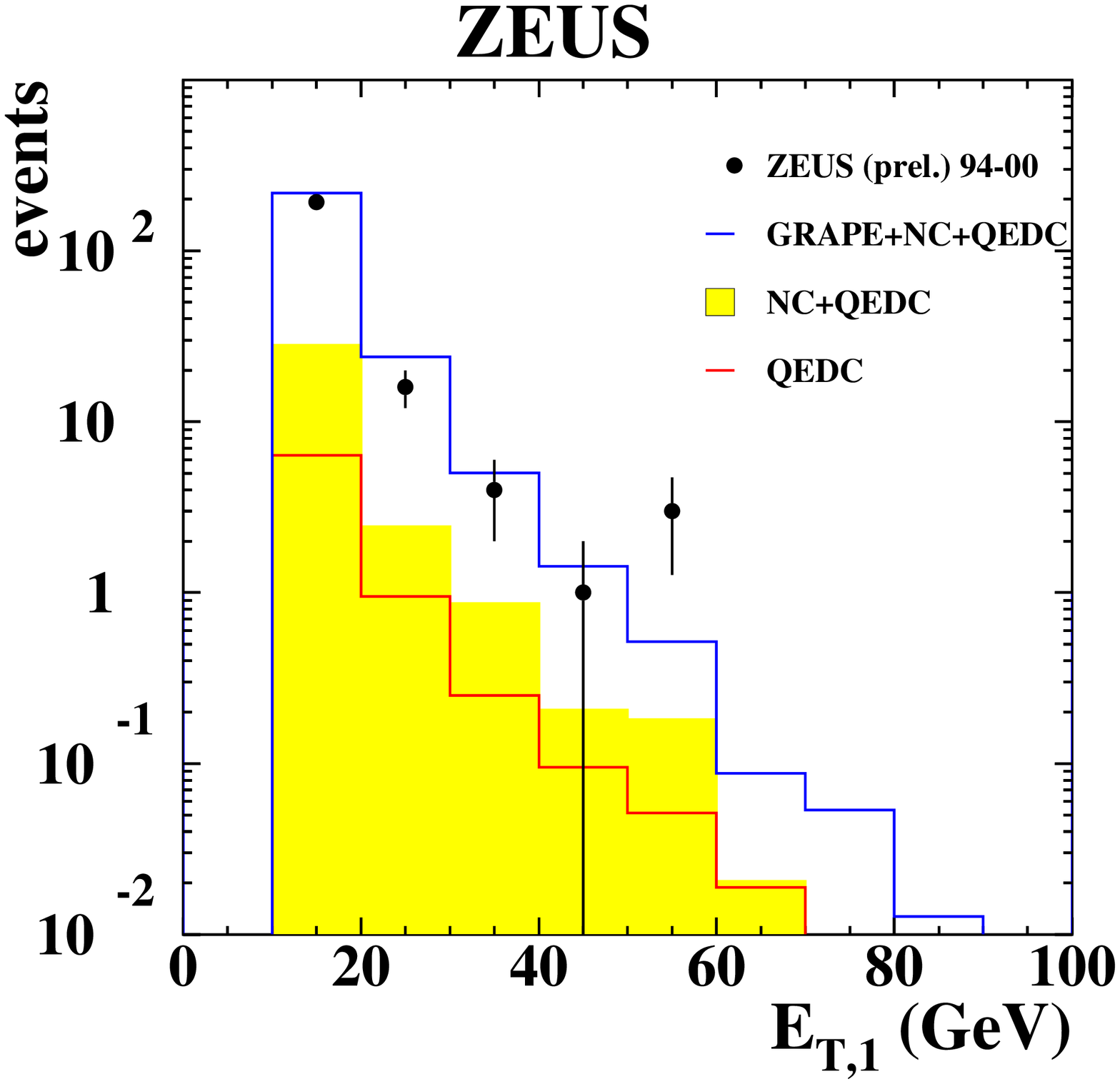,width=7.5cm}
\end{center}
\begin{picture}(0,0) \put(25,20){a)} \put(290,20){b)}  \end{picture}
\vspace{-0.5cm}
\caption{Invariant mass (a) and transverse energy ($E_T$) distribution of
the electron with the highest $E_T$ for events with two or three electrons
measured by ZEUS.
\label{fig:zeus_multi_electron}}
\end{figure}

The invariant mass distribution of events with two or
more electrons measured by ZEUS is shown in 
Fig.~\ref{fig:zeus_multi_electron}a. Also shown
is the transverse energy distribution of the electron with the
highest $E_T$. A good agreement between data and simulated SM processes
is found over the full kinematic region. 
ZEUS measures two events with $M_{1,2} > 100 \GeVx$, 
while $1.14$ events are expected. The analysed
data sample was collected in $1994$-$2000$ and
corresponds to an integrated luminosity of $130.5$ \pbinv.

The results from H1
and ZEUS are summarised in Tab.~\ref{tab:sum_multiel}.
Due to the larger angular acceptance in the backward direction,
ZEUS expects more 
$\gamma \gamma$ events, as well more background events.

\section*{Acknowledgements}
I would like to thank D. Dannheim, J. Butterworth, J. Gayler and P. Newman for the critial
reading of the manuscript.

\end{document}